\documentclass[aps,pra,reprint,amsmath,amssymb,longbibliography,superscriptaddress]{revtex4-2}

\usepackage{graphicx}
\usepackage{dcolumn}
\usepackage{bm}
\usepackage{multirow,xspace,amssymb,amsmath}

\usepackage[usenames,dvipsnames]{xcolor}
\usepackage{booktabs}
\usepackage{longtable}

\usepackage{xr} 
\makeatletter
\newcommand*{\addFileDependency}[1]{
  \typeout{(#1)}
  \@addtofilelist{#1}
  \IfFileExists{#1}{}{\typeout{No file #1.}}
}
\makeatother
\newcommand*{\myexternaldocument}[1]{
    \externaldocument{#1}
    \addFileDependency{#1.tex}
    \addFileDependency{#1.aux}
}
\myexternaldocument{res_rscan_supp_mat}

\renewcommand{\v}[1]	{\ensuremath{\mathbf{#1}}} 
\newcommand{\mr}[1]     {\ensuremath{\mathrm{#1}}}
\newcommand{\mol}[1]	{\ensuremath{_{\mathrm{#1}}}}
\newcommand{\excite}[1]    {\ensuremath{^{\mathrm{#1}}}}
\newcommand{\rppscan}   {r++SCAN}
\newcommand{\muak}      {\ensuremath{\mu_{\mr{AK}}}}
\newcommand{\tauu}{\ensuremath{\tau_{\mathrm{unif}}}~}
\newcommand{\ba}{\ensuremath{\bar{\alpha}}~}
\newcommand{\br}{\ensuremath{\mathbf{r}}}

\begin{document}

\title{Accurate and Numerically Efficient r$^2$SCAN meta-Generalized Gradient Approximation}

\author{James W. Furness}
\email{jfurness@tulane.edu}
\affiliation{Department of Physics and Engineering Physics, Tulane University, New Orleans, LA 70118}
\author{Aaron D. Kaplan}
\affiliation{Department of Physics, Temple University, Philadelphia, PA 19122}
\author{Jinliang Ning}
\affiliation{Department of Physics and Engineering Physics, Tulane University, New Orleans, LA 70118}
\author{John P. Perdew}
\affiliation{Department of Physics, Temple University, Philadelphia, PA 19122}
\affiliation{Department of Chemistry, Temple University, Philadelphia, PA 19122}
\author{Jianwei Sun}
\email{jsun@tulane.edu}
\affiliation{Department of Physics and Engineering Physics, Tulane University, New Orleans, LA 70118}

\date{\today}




\begin{abstract}
The recently proposed rSCAN functional [J. Chem. Phys. 150, 161101 (2019)] is a regularized form of the SCAN functional [Phys. Rev. Lett. 115, 036402 (2015)] that improves SCAN's numerical performance at the expense of breaking constraints known from the exact exchange-correlation functional. We construct a new meta-generalized gradient approximation by restoring exact constraint adherence to rSCAN. The resulting functional maintains rSCAN's numerical performance while restoring the transferable accuracy of SCAN.
\end{abstract}

\maketitle


{\bf Introduction } There is a fundamental trade-off at the heart of all large scale chemical and material computational studies between prediction accuracy and computational efficiency. The level of theory used must simultaneously make accurate and efficient material property predictions. For many projects, Kohn--Sham density functional theory (KS-DFT) presents an appealing compromise, delivering useful accuracy and favorable algorithmic complexity.

The Materials Project database presents a case study of finding such a balance\cite{Jain2013}, stating an ambitious mission of ``removing the guesswork from materials design by computing properties of all known materials''\cite{Persson2020}. At the time of writing, the database lists 125,000 inorganic structures calculated from KS-DFT using the Perdew-Burke-Ernzerhof (PBE) generalized gradient approximation (GGA) exchange-correlation (XC) functional \cite{Perdew1996}. While GGA functionals can be impressively accurate for many properties, they cannot be systemically accurate for all properties \cite{Perdew2008a, Ruzsinszky2009, Tran2016}, and the last 10 years have shown that meta-GGA functionals can improve predictions for similar computational cost.

Meta-GGAs are commonly designed around constraints known for the exact XC functional while minimizing the number of free parameters that must be fit. Functionals derived in this way are termed ``non-empirical,'' and we refer the reader to the supplemental material of Ref. \citenum{Sun2015} for precise definitions of all the exact constraints known for meta-GGAs. Alternatively, the functional can be built from a more flexible form that allows some exact constraints to be broken, so that free parameters can be tuned for accuracy to reference data sets. Functionals taking the latter route, termed ``empirical'' functionals, tend to be less reliable for systems outside their fitting sets, making a non-empirical functional desirable for large scale applications.

The strongly constrained and appropriately normed (SCAN) functional \cite{Sun2015} recovers all 17 exact constraints presently known for meta-GGA functionals and has shown good transferable accuracy, even for systems challenging for DFT methods. Examples include predicting accurate geometries and energetics for diverse ice and silicon phases \cite{Sun2016}, and for polymorphs of MnO\mol{2} \cite{Kitchaev2016}. SCAN accurately reproduces the complex doping driven metal-insulator transition, magnetic structure, and charge-spin stripe phases of cuprate \cite{Furness2018,Lane2018,Zhang2019a} high-temperature superconductors and iridates \cite{Lane2020}. It is one of the few functionals that predicts ice as less dense than liquid water under standard conditions \cite{Chen2017}, and its description of intermediate range van der Waals interactions has been used to study the dynamics of liquid water \cite{Chen2017,Zheng2018}. Combination of SCAN with beyond DFT techniques such as van-der-Waals functionals and the Hubbard $U$ self-interaction correction have proven effective for modelling the ionic and electronic structures of transition metal oxides\cite{Peng2017,SaiGautam2018,Zhang2019}.

Despite these successes, SCAN's utility for large scale projects is limited by its sensitivity to the density of the numerical integration grid used during calculation. This poor numerical performance in many codes mandates the use of dense integration grids which reduces SCAN's computational efficiency \cite{Yang2016,Yamamoto2019}, and divergence in the associated XC potential has hindered the generation of SCAN pseudo-potentials \cite{Yao2017,Bartok2019}. Neither limitation is inherent to the meta-GGA level or SCAN-like functionals, as we will show.

Some modifications to SCAN have been proposed to improve its accuracy for specific systems. The revSCAN functional is a simple modification to the slowly-varying limit of SCAN's correlation energy to eliminate the fourth-order term in SCAN's correlation energy density-gradient expansion \cite{Mezei2018}. The TASK functional is a complete revision of SCAN designed to accurately predict band gaps while retaining the exact constraints placed on the exchange energy \cite{Aschebrock2019}, though TASK uses a local spin-density approximation (LSDA) to model correlation. It is not expected that these modifications address the numerical inefficiencies of the parent functional.

In recent work, Bart\'ok and Yates propose a regularized SCAN termed ``rSCAN'' that aims to control SCAN's numerical challenges while changing as little as possible from the parent functional \cite{Bartok2019}. The resulting functional shows greatly improved numerical stability and enables pseudo-potential generation. While initial testing suggested that rSCAN maintained the accuracy and transferability of SCAN, expanded testing by Mej\'ia-Rodr\'iguez and Trickey \cite{Mejia-Rodriguez2019b, Bartok2019a} shows that some transferability is lost, with accuracy for atomization energies \cite{Yamamoto2020} particularly degraded.

\begin{figure}
    \centering
    \includegraphics[width=\columnwidth]{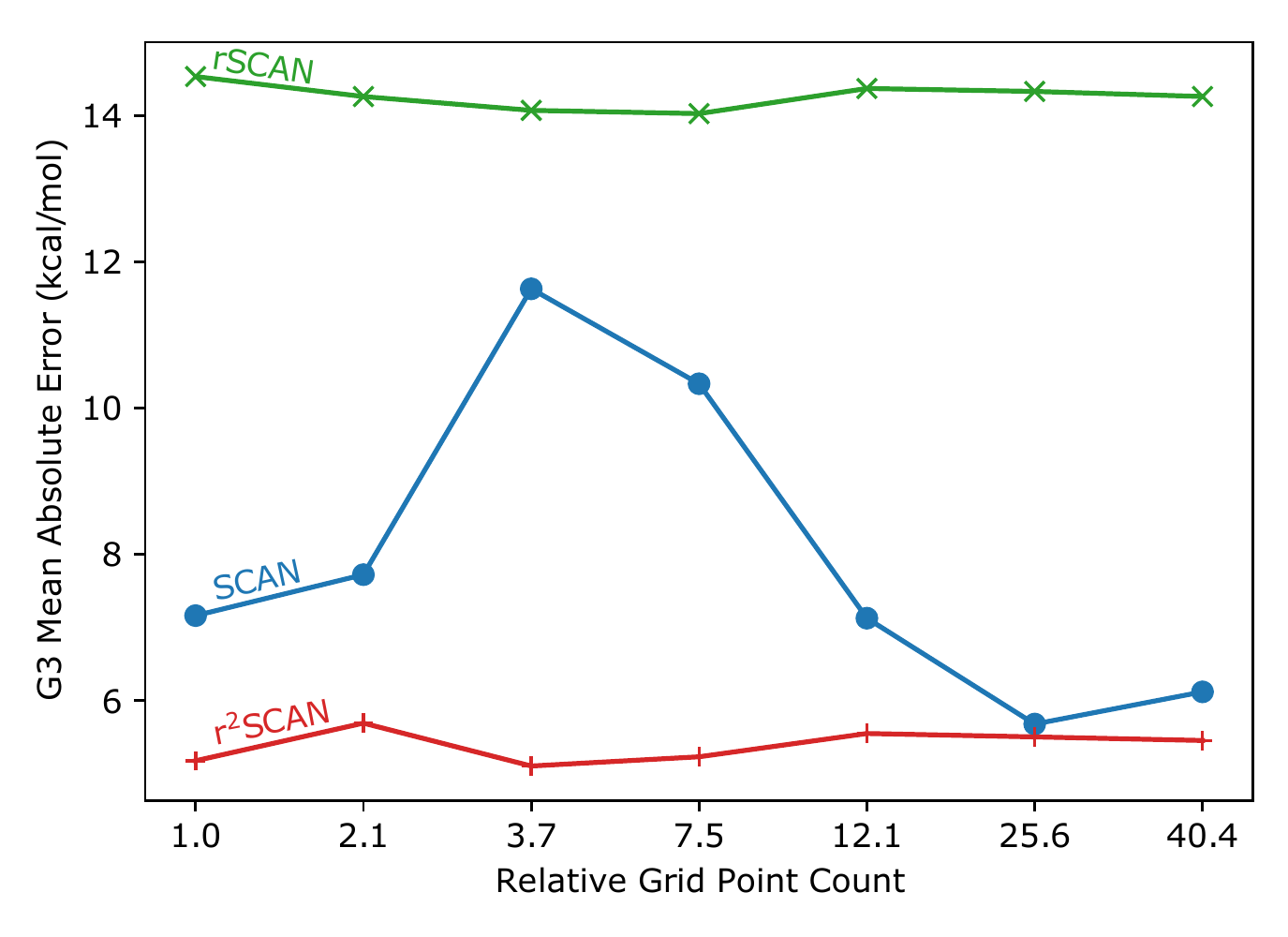}
    \caption{Mean absolute error (MAE) of atomization energies (kcal/mol) for the G3 set of 226 molecules\cite{Curtiss2000} as a function of increasing numerical integration grid density expressed relative to the smallest grid. The grids were chosen {\color{Black}from \textsc{Turbomole} \cite{Treutler1995,Balasubramani2020} grid levels 1-7}.}
    \label{fig:G3_progression}
\end{figure}

The need for a computationally-efficient revision of SCAN is made plain in Fig. \ref{fig:G3_progression}. This figure shows three meta-GGAs: SCAN and rSCAN, which have already been introduced, and a novel meta-GGA, r$^2$SCAN, that is the topic of this paper. It illustrates a grid problem that arises for SCAN in codes with localized basis functions. The horizontal axis shows increasing integration grid density, and the vertical axis shows the mean absolute error (MAE) of the G3 test set \cite{Curtiss2000} of 226 atomization energies. It would be difficult to assert that any of the grid settings present a converged SCAN energy, with SCAN errors varying unpredictably by a factor of 2. While rSCAN stabilizes SCAN numerically, its error offers little improvement over GGAs (e.g., PBE has a MAE of 22.2 kcal/mol \cite{Sun2015} for the G3 set). The need for a meta-GGA that retains the accuracy of SCAN, with the grid efficiency of rSCAN, is evident. {\color{Black} (No such grid problem is found for SCAN in the plane-wave code VASP, as shown in the Supplementary Material. However, in VASP, r$^2$SCAN seems to converge with fewer iterations than SCAN does.)}

{\bf Summary of Changes } The SCAN functional is constructed using a dimensionless kinetic energy variable,
\begin{equation}
    \alpha(\br) = \frac{\tau(\br) - \tau_W(\br)}{\tauu(\br)},
\end{equation}
where, $\tau = \sum_i |\nabla \phi_i|^2\Theta(\mu - \varepsilon_i)/2$ is the positive kinetic energy density, $\phi_i(\br)$ are the Kohn-Sham orbitals, $\Theta(\mu - \varepsilon_i)$ is the orbital occupation, $\tau_W = |\nabla n|^2/(8 n)$ is the von Weizs\"acker kinetic energy density, and $\tauu = 3(3\pi^2)^{2/3} n^{5/3}/10$ is the kinetic energy density of a uniform electron gas, $\mu$ is the chemical potential, and $\varepsilon_i$ are the orbital energies. SCAN uses $\alpha$ to tune functional performance for the local chemical environment \cite{Sun2013a}. While $\alpha$ allows SCAN to satisfy exact constraints that would be contradictory at the GGA level \cite{Sun2015a}, $\alpha$ can introduce numerical sensitivity and divergences in the XC potential \cite{Holzwarth2018, Furness2019}.

The design of rSCAN prioritizes numerical efficiency over satisfaction of exact constraints, and instead uses a regularized $\alpha^{\prime}$,
\begin{eqnarray}
    \widetilde \alpha(\br) &=& \frac{\tau(\br)-\tau_W(\br)}{\tauu(\br) + \tau_r}, \\
    \alpha^\prime(\br) &=& \frac{\widetilde \alpha(\br)^3}{\widetilde \alpha(\br)^2 + \alpha_r},
\end{eqnarray}
where $\tau_r = 10^{-4}$ and $\alpha_r = 10^{-3}$ are regularization constants. While the choice of a constant $\tau_r$ eliminates numerical instability as $\alpha \to 0$, $\alpha'$ does not retain the correct uniform and non-uniform scaling properties of $\alpha$, nor the correct uniform density limit.

For a uniform electron gas, $\alpha \to 1$, which SCAN uses to recover the LSDA exactly. In rSCAN, $\widetilde \alpha \to 1/(1+\tau_r/\tauu)$ which varies with the density, losing the correct uniform electron gas description. It has been shown that recovery of the uniform gas limit is critical for an accurate description of solids, atoms and molecules \cite{Zope2019, Bhattarai2020}.

For a slowly-varying electron gas, the exchange and correlation energies have well-known expansions in powers of the gradient of the density. Let $s = |\nabla n|/(2 k_F n)$, a dimensionless density-gradient on the scale of the Fermi wavevector $k_F = (3\pi^2 n)^{1/3}$, and $q = \nabla^2 n/(4 k_F^2 n)$ a dimensionless density-Laplacian. The gradient expansion for the exchange energy per particle $\varepsilon_x(\br)$ is \cite{Svendsen1996},
\begin{equation}
    \varepsilon_\mr{x} = \varepsilon^{\mathrm{LDA}}_\mr{x}\left[1 + \muak p + \frac{146}{2025}\left(q^2 - \frac{5}{2} p q \right) \right] + \mathcal{O}[(\nabla n)^6], \label{eq:ge4x}
\end{equation}
where $\varepsilon^{\mathrm{LDA}}_\mr{x} = -3k_F/(4\pi)$, $p = s^2$ and $\muak=10/81$. For the correlation energy, following Ref. \citenum{Perdew1996}, we define an additional dimensionless density-gradient $t = |\nabla n|/[2 k_s \phi(\zeta) n]$ on the scale of the Thomas-Fermi screening wavevector $k_s = \sqrt{4k_F/\pi}$. Here $\phi(\zeta) = [(1 + \zeta)^{2/3} + (1 - \zeta)^{2/3}]/2$ is a spin-scaling function of the spin-polarization $\zeta = (n_{\uparrow} - n_{\downarrow})/n$. Then the density-gradient expansion of the correlation energy per particle $\varepsilon_\mr{c}(\br)$ is \cite{Ma1968,Perdew1996,Sun2015},
\begin{equation}
    \varepsilon_\mr{c} = \varepsilon^{\mathrm{LSDA}}_\mr{c} + \phi(\zeta)^3 \beta(r_s) t^2,
\end{equation}
where $\beta(r_s)$ is a weakly-varying function of the Wigner-Seitz radius $r_s = (4\pi n/3)^{-1/3}$, with a maximum $\beta(0) \approx 0.066725$. The kinetic energy density $\tau$ has an analogous but unwieldy density-gradient expansion \cite{Brack1976}. It is generally understood that recovering the exact density-gradient expansion is relevant for solids \cite{Perdew2006}. These terms also affect the asymptotic behavior of $E_{\mr{xc}}$ for atoms \cite{Elliott2008}, as the asymptotic limit for atoms of large-$Z$ is a semiclassical limit that is described exactly by the LSDA at lowest-order, with the density-gradient terms modulating the higher-order terms (known accurately) \cite{Elliott2008}. Thus the uniform and slowly-varying density limits are relevant to both solid state and atomic systems.

\begin{figure}
    \centering
    \includegraphics[width=\columnwidth]{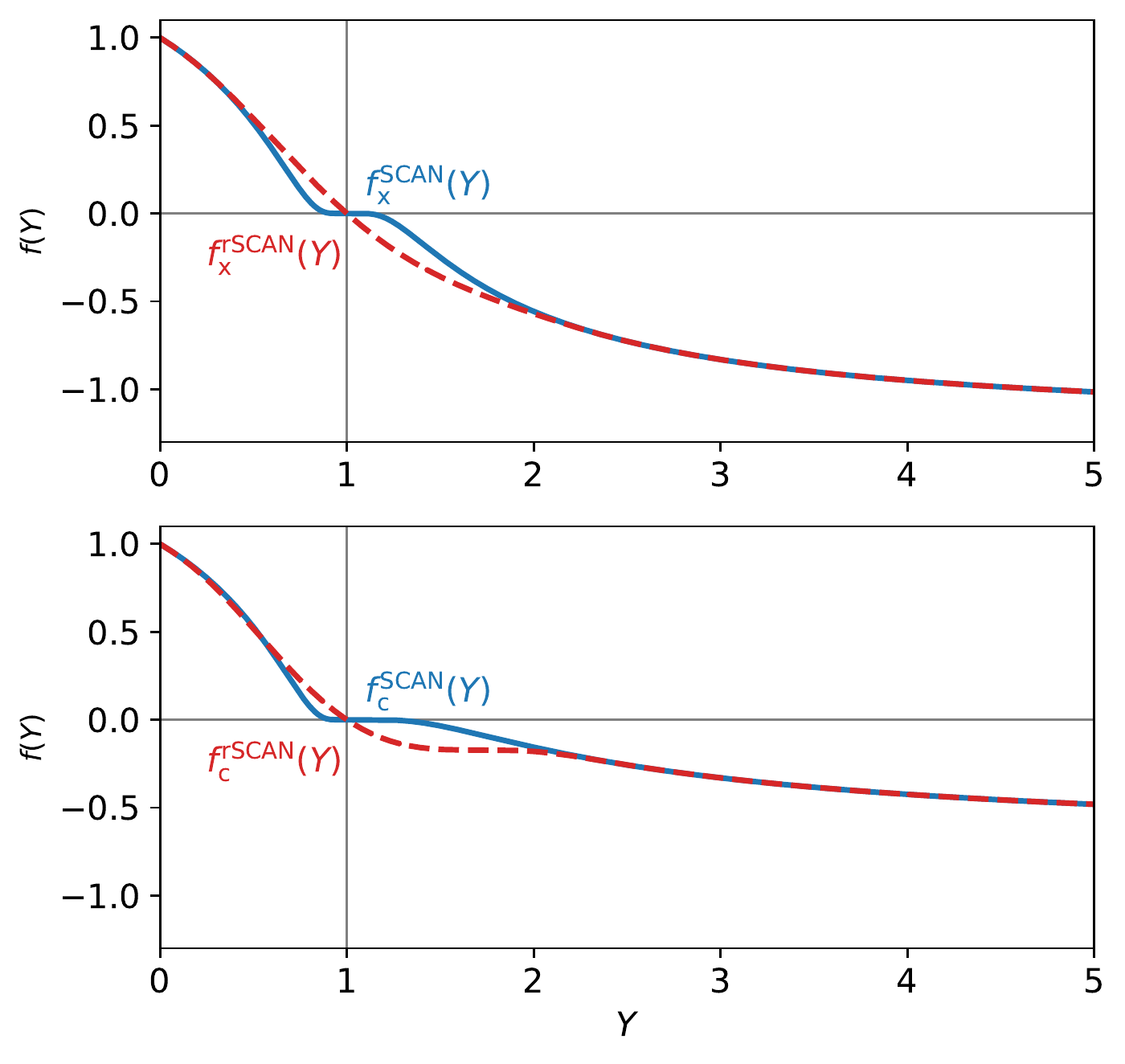}
    \caption{SCAN (blue, solid) and rSCAN (red, dashed) interpolation functions {\color{Black}plotted for a generic stand-in iso-orbital indicator ``Y'' ($\alpha$ for SCAN, $\alpha^\prime$ for rSCAN, $\ba$ for r$^2$SCAN). The functions mix $Y=0$ (single orbital) and $Y=1$ (uniform density limit for $\alpha$ and $\alpha^\prime$) energy densities.} The derivatives of the SCAN interpolation functions vanish to all orders in $Y$ at $Y \to 1$, allowing SCAN to recover the appropriate density-gradient expansions exactly in the slowly-varying limit. {\color{Black} The rSCAN interpolation functions are used with $Y = \ba$ in r$^2$SCAN and their smooth, non-vanishing first derivatives at $Y = 1$ necessitate changes from SCAN to r$^2$SCAN in the $Y = 1$ energy densities.}}
    \label{fig:ief_comparison}
\end{figure}

SCAN eliminates erroneous contributions from $\alpha$ to the second- and fourth-order slowly-varying density-gradient expansion (GE2 and GE4 respectively) of $E_{\mr{xc}}$ by using a non-analytic switching function whose value and derivatives of all orders are zero at $\alpha = 1$. Whilst theoretically convenient, constraining the interpolation function to have zero derivatives at $\alpha = 1$ results in a twisted function that harms numerical performance. The SCAN interpolation function was replaced with a smooth polynomial in rSCAN (see Fig. \ref{fig:ief_comparison}) to remove this source of numeric instability, at the expense of introducing second- and fourth-order contributions from $\alpha$ to the density-gradient expansion of $E_{\mr{xc}}$.

It is clear then that rSCAN makes wide-ranging sacrifices in exact constraint adherence in order to make a numerically-efficient meta-GGA. Here, we will show definitively that such sacrifices are needless and derive revisions to the rSCAN functional to restore exact constraint adherence without harming numerical efficiency. We apply these revisions to build a regularized-restored SCAN functional, r$^2$SCAN, which recovers the most important exact constraints of SCAN. Table \ref{tab:exact_constraint_summary} summarizes the constraint satisfaction of the functionals concerned and we stress that as only appropriate norm systems \cite{Sun2015} were used to set the free parameters, all three functionals (SCAN, rSCAN, and r$^2$SCAN) may be considered non-empirical. For brevity we only show parts of the functional that are modified in this work, and direct the reader to Section \ref{sec:eqns} of the Supplemental Materials for a full definition of the relevant equations.

\begin{table}[h]
    \centering
    \caption{Summary of exact constraint adherence for a subset of the 17 known exact constraints applicable to meta-GGA functionals. Here, GE2 denotes the second-order slowly-varying density-gradient expansion, and GE4X denotes the 4$^\mathrm{th}$ order GE for exchange.}
    \begin{tabular}{l|ccc}
                                & SCAN          & rSCAN         & r$^2$SCAN \\ \hline
        Uniform Density         & $\checkmark$  & $-$           & $\checkmark$  \\
        Coordinate Scaling      & $\checkmark$  & $-$           & $\checkmark$   \\
        GE2                     & $\checkmark$  & $-$           & $\checkmark$   \\
        GE4X                    & $\checkmark$  & $-$           & $-$            \\
    \end{tabular}
    \label{tab:exact_constraint_summary}
\end{table}

{\color{Black} There are many situations where the exact exchange-correlation potential and energy density can be expected to be reasonably smooth (see, e.g., the plots of highly accurate exchange-correlation potentials and energy densities of simple hydrides in Ref. \citenum{Gritsenko1996}). In general, the exact Kohn-Sham exchange-correlation potential need not be smooth, as demonstrated by the Perdew-Parr-Levy-Balduz theorem \cite{Perdew1982}: the exchange-correlation potential, as a function of the number of electrons $N$, exhibits discontinuities across integer values of $N$, with steps and peaks in the low-density region between two separated dissimilar systems. However, a semilocal functional cannot recover the precise behaviors of the exact exchange-correlation energy and potential, and instead averages over them. Therefore we consider smoothness of the energy density and potential to be a necessary construction principle of semilocal approximate density functionals. A construction principle is any physically- or mathematically-motivated principle that can supplement the design of a first-principles density functional approximation. }

The correct uniform- and non-uniform- scaling properties of $\alpha$, as well as the correct uniform density limit of $E_\mr{xc}$, {\color{Black} are recovered in r$^2$SCAN} by regularizing $\alpha$ as
\begin{equation}
    \ba = \frac{\tau - \tau_W}{\tauu + \eta \tau_W},
\end{equation}
where $\eta = 10^{-3}$ is a simple regularization parameter. Note that because $\tau \geq \tau_W$, $\ba$ has the same range as $\alpha$, $0 \leq \ba < \infty$. This is distinct from the dimensionless kinetic energy variable suggested by Ref. \citenum{Furness2019},
\begin{equation}
    \beta = \frac{\tau - \tau_W}{\tau + \tauu},
\end{equation}
which ranges between $0 \leq \beta < 1$, and has less rapidly-varying derivatives than $\alpha$. As this work seeks revisions to SCAN, we will not consider $\beta$ or related iso-orbital indicators here {\color{Black}and adopt $\ba$ as the iso-orbital indicator used throughout r$^2$SCAN}.

{\color{Black}SCAN uses the iso-orbital indicator variable $\alpha$ to drive interpolation functions $f_\mr{x/c}(\alpha)$ for exchange and correlation. The rSCAN functional replaces the original SCAN interpolation functions with a polynomial function of $\alpha^\prime$ when $\alpha^\prime < 2.5$ that smooths out the plateau-like behavior of the original near $\alpha = 1$. The r$^2$SCAN functional adopts the rSCAN interpolation function but uses $\bar{\alpha}$ as the indicator variable. Both the SCAN and rSCAN interpolation functions are shown in Figure \ref{fig:ief_comparison} as functions of a generic indicator.

The SCAN interpolation function was designed to have vanishing derivatives at $\alpha=1$, but the rSCAN replacements go linearly through zero at this point. As a result of these non-vanishing derivatives, the interpolation function makes spurious contributions to the slowly-varying density-gradient expansions that break the corresponding exact constraints. The r$^2$SCAN functional recovers the gradient expansions through $\mathcal{O}[(\nabla n)^2]$ while using the rSCAN polynomial interpolation by directly canceling spurious terms in the slowly-varying energy densities.

For exchange we recover the gradient expansion by replacing the $x(p, \alpha^\prime)$ function of SCAN and rSCAN with,
\begin{equation}
    x(p) = \{C_{\eta}C_{2\mr{x}}\exp[-p^2/d_{p2}^4] + \muak \}p.\label{eq:newx}
\end{equation}
Here, the constants $C_{\eta} = 20/27 + 5\eta/3$, depending on the regularization parameter $\eta = 10^{-3}$, and $C_{2x} \approx -0.162742$ eliminate erroneous contributions from $df_\mr{x}(\ba)/d\ba$ at $\ba \to 1$, and $d_{p2} = 0.316$ is a damping parameter determined as the maximal value (therefore least damped) required to recover SCAN's error for the rare gas atom and jellium surface appropriate norms described in Ref. \citenum{Sun2015}. Replacing $\alpha^\prime$ with $\ba$  and $x(p, \alpha^\prime)$ with Eq. \ref{eq:newx} in the rSCAN functional defines r$^2$SCAN exchange, and restores the uniform density limit, correct scaling properties, and GE2 for exchange (GE2X).} As in Ref. \citenum{Sun2015} and earlier work, we employ the exact spin-scaling equality for the exchange energy \cite{Oliver1979}, thus only formulas for spin-unpolarized exchange need to be displayed.

From Eq. \ref{eq:ge4x}, we see that a meta-GGA recovering GE4X must either explicitly include the Laplacian of the density as an ingredient, or recover $q$-dependent terms via integration by parts on $\tau$. The latter method, used in SCAN, is theoretically sound but likely introduces further numerical instabilities due to an increased sensitivity to $\alpha$. Furthermore, the gradient expansion for the correlation energy is known only to second order, and the relevance of GE4X (beyond GE2X terms) to real systems has not been established. To ensure that our functional is numerically stable we only consider GE2X here and defer further discussion of GE4X and its difficulties to a further publication in the near future.

{\color{Black}The gradient expansion for correlation is only known to second order, and we recover it by replacing the $g(A t^2)$ function which appears in the slowly-varying correlation energy density of rSCAN and SCAN with,}
\begin{widetext}
\begin{align}
    g(A t^2,\Delta y) &= [1 + 4(A t^2 - \Delta y) ]^{-1/4}, \label{eq:gt} \\
    \Delta y &= \frac{\Delta f_{c2}}{27 \gamma d_s \phi^3 w_1} \left\{20r_s\left[\frac{\partial \varepsilon_c^{\mr{LSDA0}}}{\partial r_s} - \frac{\partial \varepsilon_c^{\mr{LSDA1}}}{\partial r_s} \right] - 45\eta[\varepsilon_c^{\mr{LSDA0}} - \varepsilon_c^{\mr{LSDA1}}] \right\} p \exp[-p^2/d_{p2}^4]. \label{eq:newgt}
\end{align}
\end{widetext}
{\color{Black}Here $\Delta y$ is the new term introduced in r$^2$SCAN that eliminates erroneous contributions from $df_\mr{c}(\ba)/d\ba$ at $\ba \to 1$. Replacing $\alpha^\prime$ with $\ba$ and $g(A t^2)$ with Eqs. \ref{eq:gt} and \ref{eq:newgt} in rSCAN correlation defines r$^2$SCAN correlation and approximately recovers GE2C.} In Eq. \ref{eq:newgt}, $\Delta f_{c2} \approx -0.711402$ and $\gamma=(1-\ln2)/\pi^2\approx 0.031091$ are constants; all other quantities are functions defined in the Supplemental Materials. The correlation gradient expansion of r$^2$SCAN becomes exact whenever $|\nabla \zeta| = 0$, and approximately recovers GE2C for all other values of $\nabla \zeta$. Thus r$^2$SCAN recovers GE2C exactly for spin-unpolarized systems, where the correlation energy is likely most negative, and for fully spin-polarized systems, where the correlation energy is likely least negative. Between these limits, the r$^2$SCAN correlation gradient expansion is a reasonable approximation to the true gradient expansion.

\begin{figure*}[t]
    \centering
    \includegraphics[width=\textwidth]{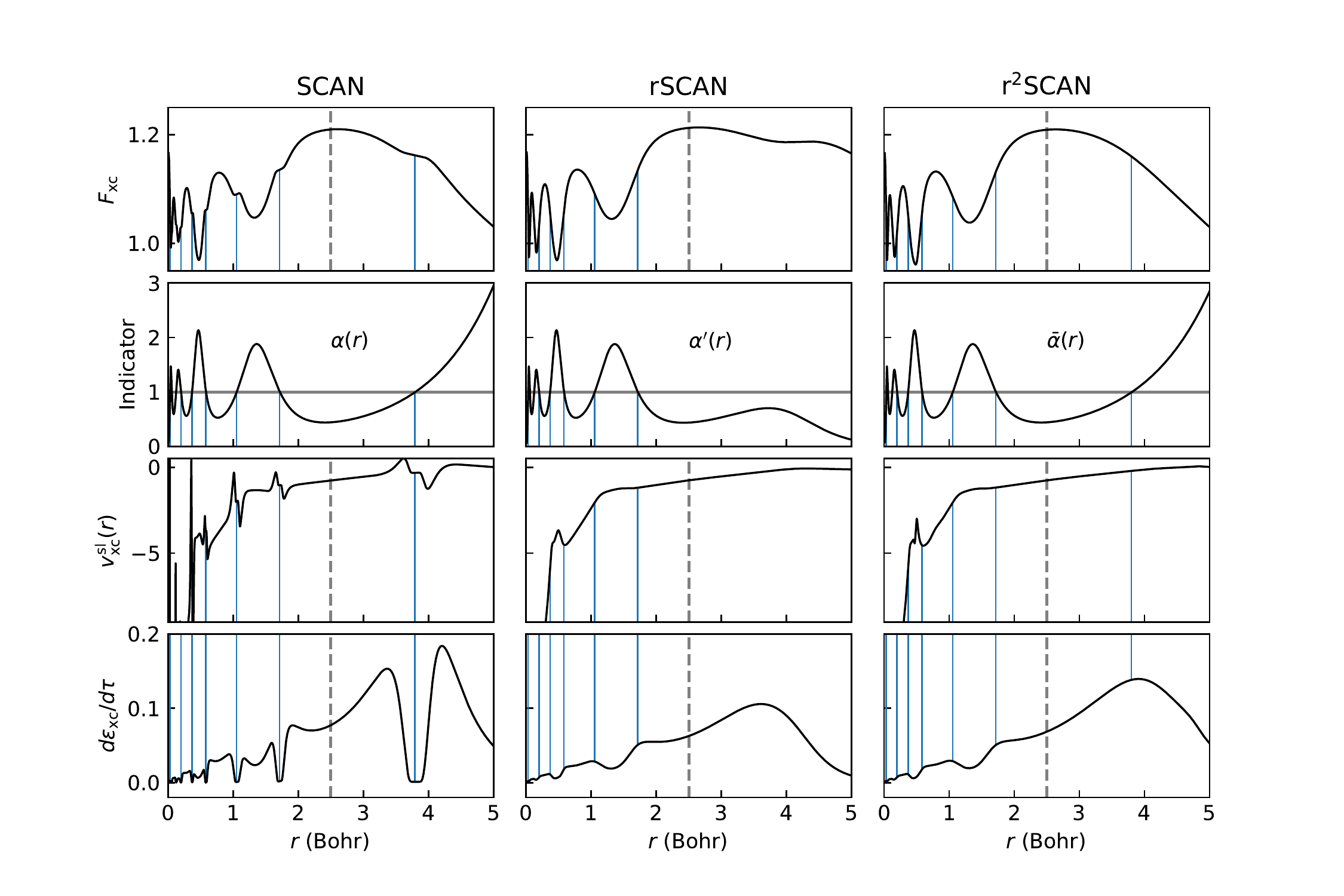}
    \caption{(Top) Exchange-correlation enhancement factors, (middle-upper) iso-orbital indicator $\alpha(r)$, $\alpha^\prime(r)$, or $\ba(r)$ as appropriate, (middle-lower) semi-local part of the exchange-correlation potential as in Eq. \ref{eq:vxcsl}, and (bottom) derivative of exchange-correlation energy density with respect to kinetic energy density. Calculated for the xenon atom from accurate Hartree--Fock Slater orbitals \cite{Clementi1974} for the SCAN \cite{Sun2015}, rSCAN \cite{Bartok2019} and r$^2$SCAN functionals. The VASP \cite{Kresse1993,Kresse1994,Kresse1996,Kresse1996a} projector-augmented wave \cite{Joubert1999} pseudopotential cutoff radius (2.5 Bohr) is illustrated by a dashed vertical line. Solid vertical lines show where $\alpha=1$.}
    \label{fig:xe_pot}
\end{figure*}

{\bf Enhancement Factors and Derivatives} These modifications should not degrade the good numerical performance of rSCAN. Their effect can be seen in Figure \ref{fig:xe_pot} which compares the XC enhancement factor and XC potential components of SCAN and r$^2$SCAN for the xenon atom. The implicit orbital dependence of $\tau$ dependent meta-GGA functionals prevents direct evaluation of a multiplicative KS potential, and such functionals are more commonly implemented using derivatives with respect to individual orbitals in a generalized Kohn--Sham scheme \cite{Seidl1996, Neumann1996, Adamo2000}. Let $\epsilon_{xc} = n \varepsilon_{xc}$ be the XC energy density. We can identify a multiplicative component of the potential,
\begin{equation}
    v_\mathrm{xc}^\mathrm{sl}(\mathbf{r}) = \frac{\partial \epsilon_\mr{xc}(\mathbf{r})}{\partial n(\mathbf{r})} - \nabla \cdot \left[ \frac{\partial \epsilon_\mr{xc}(\mathbf{r})}{\partial |\nabla n(\mathbf{r})|} \frac{\nabla n(\mathbf{r})}{|\nabla n(\mathbf{r})|} \right], \label{eq:vxcsl}
\end{equation}
and summarize the non-multiplicative component as the derivative of the energy density with respect to $\tau$. Both are shown in Figure \ref{fig:xe_pot}.

While the overall inter-shell features of the $F_\mr{xc}$ are similar between the two functionals, SCAN shows plateaus where $\alpha= 1$, while r$^2$SCAN is smooth throughout. This behavior is echoed as sharp oscillations in both the multiplicative potential component and non-multiplicative $\tau$ derivative for SCAN, which contrasts with the smooth equivalents for r$^2$SCAN. We suggest that r$^2$SCAN may make generation of meta-GGA pseudopotentials feasible. {\color{Black}Note that on the scale of Figure \ref{fig:xe_pot} both $\alpha$ and $\bar{\alpha}$ appear to diverge. While it is true that $\alpha$ diverges, the $\eta$ regularization parameter ensures $\bar{\alpha}$ remains finite, with a final maxima occurring around 8 Bohr after which it asymptotically approaches 0. Larger values of $\eta$ cause the asymptotic return to occur closer to the nuclei, but cause the regularization to have a greater impact on predicted energies. We defer detailed discussion of $\eta$ and single orbital system potential divergence to Supplemental Material Section S3.}

{\bf Atomization Energies} Reference \citenum{Mejia-Rodriguez2019b} shows that atomization energies are particularly problematic for rSCAN, with its error being roughly twice that of SCAN's. As such, we take the G3 set of 226 atomization energies \cite{Curtiss2000} as a primary means of assessing the effect of constraint restoration. Table \ref{tab:all_test_stats} summarizes the accuracy of the functionals for the G3 set using the most dense grid available. We find that the error for rSCAN is roughly twice that of SCAN, consistent with other studies. The new r$^2$SCAN functional shows similar accuracy to SCAN, supporting the importance of exact constraint adherence.

We restate that the improved numerical efficiency of r$^2$SCAN is immediately apparent from Figure \ref{fig:G3_progression}, which shows accuracy for the G3 test set as a function of integration grid density. r$^2$SCAN shows consistent error with grid density, similar to rSCAN and in sharp contrast to SCAN. This figure should stand as a stark warning that studies comparing total energies from SCAN must carefully test for grid convergence, and shows the utility of the new regularized-restored functional which achieves consistently-good accuracy with even the smallest grids.

{\bf Further Testing} The transferablility of the new functionals was further tested for 76 reaction barrier heights \cite{Zhao2005}, 22 weak interaction energies \cite{Jurecka2006}, and 20 lattice constants \cite{Sun2011}, with error statistics summarized in Table \ref{tab:all_test_stats}. All functionals show similar performance across the test sets, with the exception of the G3 atomization energy set as discussed above. The LC20 set was assessed by fitting the stabilized jellium equation of state (SJEOS) \cite{Staroverov2004,Alchagirov2001} to single point energies at a range of lattice volumes around equilibrium.

\begin{table}[h]
    \centering
    \caption{Mean error (ME) and mean absolute error (MAE) of {\color{Black}TPSS}\cite{Tao2003}, SCAN\cite{Sun2015}, rSCAN\cite{Bartok2019}, and r$^2$SCAN for the the G3 set of 226 molecular atomization energies \cite{Curtiss2000}, the BH76 set of 76 chemical barrier heights \cite{Zhao2005}, the S22 set of  22 interaction energies between closed shell complexes \cite{Jurecka2006}, and the LC20 set of 20 solid lattice constants \cite{Sun2011}. Errors for G3, BH76, and S22 sets are in kcal/mol while errors for LC20 are in \AA{}. We did not make corrections for basis set super-position error for the S22 set which used the aug-cc-pVTZ basis set\cite{Dunning1989}. All calculations for G3 and BH76 used the 6-311++G(3df,3pd) basis set \cite{Clark1983,Curtiss2000}. Details of the computational methods are included in Section \ref{sec:comp_metd} of the Supplemental Materials.}
    \begin{tabular}{c|rr|rr|rr|rr}
            & \multicolumn{2}{c|}{G3}  & \multicolumn{2}{c|}{BH76} & \multicolumn{2}{c|}{S22} & \multicolumn{2}{c}{LC20} \\
            & \multicolumn{1}{c}{ME} & \multicolumn{1}{c|}{MAE} & \multicolumn{1}{c}{ME} & \multicolumn{1}{c|}{MAE} & \multicolumn{1}{c}{ME} & \multicolumn{1}{c|}{MAE} & \multicolumn{1}{c}{ME} & \multicolumn{1}{c}{MAE} \\
        \hline
        \color{Black}TPSS & -5.2 & 5.8 & -8.6 & 8.6 & -3.4 & 3.4 & 0.033 & 0.041\\
        SCAN & -5.0 & 6.1 & -7.7 & 7.7 & -0.5 & 0.8 & 0.009 & 0.015 \\
        rSCAN & -14.0 & 14.3 & -7.4 & 7.4 & -1.2 & 1.3 & 0.020 & 0.025 \\
        r$^2$SCAN & -4.5 & 5.5 & -7.1 & 7.2 & -0.9 & 1.1 & 0.022 & 0.027 \\
    \end{tabular}

    \label{tab:all_test_stats}
\end{table}

{\bf Conclusions} {\color{Black} We recently learned of a ``de-orbitalization'' \cite{Smiga2015,Smiga2017a,Mejia-Rodriguez2017} of our r$^2$SCAN that replaces the exact orbital dependent Kohn-Sham kinetic energy density $\tau$ by an posited function of $n$, $\nabla n$, and $\nabla^2 n$, called ``r$^2$SCAN-L''\cite{Mejia-Rodriguez2020}. This speeds up computations while somewhat reducing overall accuracy, though interestingly the accuracy of the magnetic moment of metallic Fe is restored by the de-orbitalization to the good level of LSDA and PBE. As a possible explanation, we note that the exact $\tau$ has a fully nonlocal dependence upon the electron density $n$ that is needed to satisfy some exact constraints and is probably needed for optimal accuracy in atoms, molecules, and insulators. This full nonlocality may however be somewhat harmful for metals, where metallic screening can favor truly semi-local approximations to the valence-valence exchange-correlation energy.}

We have presented r$^2$SCAN as a functional combining SCAN's transferable accuracy from exact constraint satisfaction with rSCAN's numerical efficiency. r$^2$SCAN satisfies the most important exact constraints of SCAN. In our future work, we will assess the importance of the GE4X terms beyond GE2X (recovered by SCAN, but not by r$^2$SCAN) as an exact constraint. We draw this conclusion from the competitive accuracy shown for the diverse test sets of Table \ref{tab:all_test_stats} and the rapid grid convergence of Figure \ref{fig:G3_progression}. The XC potential analysis from Figure \ref{fig:xe_pot} suggests that r$^2$SCAN will be preferable when a smooth potential is critical. {\color{Black}In particular, it should now be more practical to construct a pseudopotential, and to evaluate the second functional derivative which can play the role of an exchange-correlation kernel in time-dependent density functional applications.} We expect the new regularized-restored SCAN functional to bridge the gap between accuracy and numerical efficiency and enable meta-GGA use in large-scale computational studies.

\section*{Supporting Information}
{\color{Black} S1: Computational Methods and grid convergence in VASP, S2: r$^2$SCAN Working Equations, S3: Determining $\eta$ regularization parameter and XC potential divergence, S4: reference atomic energies, S5: full test set data.}

\begin{acknowledgements}
J.W.F., J.N., and J.S. acknowledge the support of the U.S. DOE, Office of Science, Basic Energy Sciences Grant No. DE-SC0019350 (core research). A.D.K. and J.P.P. acknowledge the support of the U.S. Department of Energy, Office of Science, Basic Energy Sciences, through Grant No. DE-SC0012575 to the Energy Frontier Research Center: Center for Complex Materials from First Principles. We thank Albert Bart\'ok-Partay, and Daniel Mejia-Rodriguez for their invaluable discussions around the ideas presented here. J.P.P. thanks Natalie Holzwarth for pointing out that the SCAN exchange-correlation potential for an atom diverges in the tail of the density, making pseudo-potential construction difficult.
\end{acknowledgements}

\bibliography{r2SCAN_Paper}

\clearpage
\renewcommand{\thepage}{S\arabic{page}}
\renewcommand{\thesection}{S\arabic{section}}
\renewcommand{\theequation}{S\arabic{equation}}
\renewcommand{\thetable}{S\arabic{table}}
\renewcommand{\thefigure}{S\arabic{figure}}
\renewcommand{\epsilon}{\varepsilon}

\renewcommand{\v}[1]	{\ensuremath{\mathbf{#1}}} 
\renewcommand{\mr}[1]     {\ensuremath{\mathrm{#1}}}
\renewcommand{\mol}[1]	{\ensuremath{_{\text{#1}}}}
\renewcommand{\excite}[1]    {\ensuremath{^{\text{#1}}}}
\renewcommand{\rppscan}   {r++SCAN}
\renewcommand{\muak}      {\ensuremath{\mu_{\mr{AK}}}}
\renewcommand{\tauu}{\ensuremath{\tau_{\text{unif}}}\xspace}
\renewcommand{\ba}{\ensuremath{\bar{\alpha}}\xspace}
\renewcommand{\br}{\ensuremath{\bm{r}}\xspace}


\setcounter{section}{0}
\setcounter{page}{1}
\setcounter{equation}{0}
\setcounter{table}{0}
\setcounter{figure}{0}

\onecolumngrid

\section*{Supplemental Materials for: ``Accurate and Numerically Efficient r$^2$SCAN meta-Generalized Gradient Approximation''}




\section*{Contents}

Section \ref{sec:comp_metd} provides essential input parameters needed to reproduce the calculations reported in the main text. Section \ref{sec:eqns} provides the equations of r$^2$SCAN, with numeric parameters given. Section \ref{sec:ref_at_en} tabulates reference atomic energies for the nitrogen and neon atoms. Section \ref{sec:raw_dat} tabulates all resultant data for the test sets in the main text. All data referenced in the main text is made available here. Machine-readable data will be made available at reasonable request.

Tables \ref{tab:G3_SCAN}, \ref{tab:G3_rSCAN}, and \ref{tab:G3_r2SCAN} list all errors of SCAN, rSCAN, and r$^2$SCAN, respectively for the G3 set. Table \ref{tab:full_BH76} lists all errors for the BH76 set; Table \ref{tab:full_S22} does this for the S22 set, and Table \ref{tab:full_LC20} for the LC20 set.

\section{Computational Methods \label{sec:comp_metd}}

Within \textsc{Turbomole} \cite{Treutler1995,Balasubramani2020}, all calculations used the following parameters:
\begin{itemize}
    \item Gaussian basis set 6-311++G(3df,3pd), excepting the S22 set, which used aug-cc-pVTZ. \cite{Clark1983,Curtiss2000}
    \item self-consistent convergence at 10$^{-6}$ eV.
\end{itemize}
The G3 set of atomisation energies \cite{Curtiss2000}, the BH76 set of reaction barrier heights \cite{Zhao2005}, and the S22 set of weakly-interacting molecular complex formation energies \cite{Jurecka2006} were calculated in \textsc{Turbomole}. The S22 calculation included no counterpoise correction, however it is evident from Table \ref{tab:full_S22} that the scores are similar to that found in Ref. \cite{Sun2015}.

Within VASP \cite{Kresse1993,Kresse1994,Kresse1996,Kresse1996a}, all calculations presented used the following parameters:
\begin{itemize}
    \item ISMEAR function (controlling partial occupancy near the Fermi level) set to -5 (tetrahedron method with Bl\"ochl corrections), except for Sr, where it was set to 1 (Methfessel--Paxton scheme)
    \item ENCUT set to 650 eV for plane wave cutoff energy (except for Na, where it was set to 1000 eV)
    \item $\Gamma$-centered grid in $\bm k$-space, with KSPACING set to 0.1 \AA{}$^{-1}$
    \item ADDGRID set to true (optional finer grid to calculate augmentation charge density)
    \item PREC set to ``Accurate''.
\end{itemize}
{\color{Black} The LC20 set of cubic lattice constants \cite{Sun2011} was calculated in VASP. The ENCUT parameter in VASP controls the number of reciprocal space vectors included in the basis set. For $\bm{G}$ a reciprocal lattice vector and $\bm{k}$ the crystal momentum, VASP includes all reciprocal space vectors $|\bm{G} + \bm{k}|^2/2 < \mathrm{ENCUT}$. We have selected values of ENCUT that are much larger than the minimal value required by the pseudopotential (these range between 119.552 eV for Al to 487.335 eV for F).

Note that the PBE pseudopotential for Na required a much larger cutoff (645.64 eV when including \textit{s} semi-core states) than the corresponding GW pseudopotential (259.494 eV, also including \textit{s} semi-core states). The GW pseudopotential for Na was used here. To ensure that our calculations could be adequately reproduced regardless of the pseudopotential parameters, we have set the Na cutoff much higher than the other elements.

VASP does not use basis sets for real-space integration; the fineness of the real-space mesh and the reciprocal-space mesh are controlled by the KSPACING parameter. Smaller values of KSPACING increase the fineness of both meshes, generally improving the quality of numeric integration, and allowing for more $\bm{k}$-states to be included in the Hamiltonian. We have selected a value of 0.1 \AA{}$^{-1}$ much smaller than the recommended default, 0.5 \AA{}$^{-1}$.

We have also performed a convergence test of the mean absolute error (MAE) for AE6 test set \cite{Lynch2004} of six atomization energies (a subset of G3) with respect to the real-space grid in VASP. The size of the real-space grid in VASP is controlled by the precision (PREC, set to ``accurate'' here) and the energy cutoff (ENCUT). One may also specify the number of grid points per axis using the NGX, NGY, and NGZ tags. For this test, the ``fine'' grid settings NGXF, etc. were left to be automatically determined by the combination of PREC, ENCUT, and the grid points per axis. A default grid was selected by leaving ENCUT as the default from the pseudopotentials, and the grid points per axis unspecified. As is common for molecular and atomic calculations in VASP, only the $\Gamma$ point was used to sample $\bm{k}$ space. In Table \ref{tab:vasp_ae6_grid_conv}, it can be seen that the default grid setting is completely sufficient to resolve the SCAN potential.}

\begin{longtable}{c|ccccccc}
    \caption{ Real-space grid convergence of the AE6 test set \cite{Lynch2004} of six atomization energies in VASP. Mean absolute errors are given in kcal/mol. For each setting besides ``Default'', ENCUT was set to 1000 eV, and the number of grid points along each axis NGX=NGY=NGZ was set to the displayed number of points per axis. For the ``Default'' grid setting, ENCUT was left unspecified, as were NGX, NGY, and NGZ. In all cases, only the $\Gamma$ point was used for $\bm{k}$-space sampling. The authors were recently made aware of a long-standing bug affecting convergence of spin-polarized calculations (ISPIN=2) using meta-GGAs in VASP. The routine used here was updated to correctly treat spin-polarization. \label{tab:vasp_ae6_grid_conv}}\\
    \toprule
    Points per axis & Default & 60 & 80 & 100 & 120 & 140 & 150 \\ \midrule
    \endhead
    \bottomrule
    \endfoot
    SCAN & 3.581 & 4.018 & 3.951 & 3.906 & 4.032 & 3.989 & 3.989 \\
    rSCAN & 7.054 & 6.958 & 7.032 & 7.026 & 7.031 & 7.033 & 7.027 \\
    r$^2$SCAN & 3.491 & 3.866 & 3.767 & 3.765 & 3.765 & 3.765 & 3.765 \\
\end{longtable}

\clearpage

\section{\lowercase{r}$^2$SCAN Equations\label{sec:eqns}}

The full equations required for implementing r$^2$SCAN are given below. Note that by construction $\bar\alpha \ge 0$ (also $\alpha$ and $\alpha\prime$). In pseudopotential codes (e.g., VASP) or through rounding errors in regions of very small density, $\bar\alpha$ (or $\alpha$) can become negative, which can cause numerical problems for interpolation functions that do not consider this possibility. An additional condition was included to Eqs. \ref{eq:fx} and \ref{eq:fc} to consistently handle negative $\bar\alpha$ regions. These provisions were essential to reliably converge calculations in VASP, as neither extending the polynomial interpolation nor setting a constant $f(\bar\alpha < 0) = 1$ were sufficient.

The ultimate choice of regularization for the unphysical region $\ba < 0$, as shown in Eqs. \ref{eq:fx} and \ref{eq:fc}, is consistent with the current implementation of SCAN in VASP. The current SCAN subroutines in VASP simply extend SCAN's interpolating functions, $f_x(\alpha)$ and $f_c(\alpha)$, to $\alpha < 0$; our implementation does the same. We do not ascribe a physical interpretation to these regions, nor to the numeric regularization chosen. They are necessary computational artifices.

An exchange functional for a density with arbitrary spin polarization, $n_{\uparrow}$ and $n_{\downarrow}$, can be constructed from a spin-unpolarized functional via the spin-scaling relation \cite{Oliver1979}
\begin{equation}
    E_{\mr{x}}[n_{\uparrow},n_{\downarrow}] = \frac{1}{2}\left\{E_{\mr{x}}[2n_{\uparrow}]+E_{\mr{x}}[2n_{\downarrow}] \right\}.
\end{equation}
Thus we present equations only for spin-unpolarized exchange energy functionals.

Any constants not defined here nor in the main text are unchanged from rSCAN.

\subsection{Exchange}
 ~
\begin{align}
     E_{\mr{x}}^\mathrm{r^2SCAN}[n] &= \int n(\br) \epsilon_\mr{x}^\mathrm{r^2SCAN}(\br) d^3 r \\
    \epsilon_\mr{x}^\mathrm{r^2SCAN} &= \epsilon_\mr{x}^{\mr{LDA}}(r_s)F_\mr{x}^\mathrm{r^2SCAN}(p, \bar\alpha)\\
    \epsilon_\mr{x}^{\mr{LDA}}(r_s) &= -\frac{\frac{3}{4\pi}\left(\frac{9\pi}{4}\right)^{1/3}}{r_s}\\
	F_\mathrm{x}^\mathrm{r^2SCAN}(p, \bar\alpha) &= \left\{h^1_\mathrm{x}(p) + f_\mr{x}(\bar{\alpha})\left [h^0_\mathrm{x} - h^1_\mathrm{x}(p)\right ]\right\} g_\mathrm{x}(p)\\
	\bar{\alpha}(p, \alpha) &= \frac{\alpha}{1 + \eta \frac{5}{3}p} = \frac{\tau - \tau_{\mathrm{W}}}{\tau_{\mathrm{U}} + \eta\tau_{\mathrm{W}}} \\
    f_\mr{x}(\bar\alpha) &= \begin{cases}
        \exp\left[-\frac{c_{1\mr{x}}^{\mr{SCAN}}\bar\alpha}{1 - \bar\alpha}\right] & \bar{\alpha} < 0 \\
        \sum_{i=0}^7c_{\mr{x}i}\bar\alpha^i & 0 \le \bar\alpha \leq 2.5 \\
        -d_{\mr{x}}^{\mr{SCAN}}\exp\left[\frac{c_{\mr{2x}}^{\mr{SCAN}}}{1-\bar\alpha}\right] & \bar\alpha > 2.5
        \end{cases}\label{eq:fx}\\
	h^0_\mathrm{x} &= 1 + k_0\\
	h^1_\mathrm{x}(p) &= 1 + k_1 - \frac{k_1}{1 + \frac{x(p)}{k1}}\\
	x(p) &= \left(C_\eta C_2\exp[-p^2/d_{p2}^4] + \mu\right)p\\
	C_\eta &= \left[\frac{20}{27} + \eta\frac{5}{3}\right]\\
	C_2 &= -\sum_{i=1}^{7}ic_{\mr{x}i}[1 - h^0_{\mr{x}}] \approx -0.162742 \\
	g_\mathrm{x}(p) &= 1 - \exp\left[\frac{-a^{\text{SCAN}}_1}{p^{1/4}}\right]
\end{align}

The values of the new constants appearing in the previous equations are $\eta = 0.001$ and $d_{p2}=0.361$.

Some constants appearing here are unchanged from either SCAN or rSCAN. Note that for the rSCAN interpolating polynomial we take the convention of indexing the coefficient by the power of $\alpha$ it multiplies, thus the vector $\bm{c}_x$ begins at $c_0$. These are:
\begin{align}
    \bm{c}_{x} &= (1, -0.667, -0.4445555, -0.663086601049, 1.451297044490,\nonumber \\
    & -0.887998041597, 0.234528941479, -0.023185843322) \\
    c^{\mr{SCAN}}_{1x} &= 0.667 \\
    c^{\mr{SCAN}}_{2x} &= 0.8 \\
    d^{\mr{SCAN}}_{x} &= 1.24 \\
    k_0 &= 0.174 \\
    k_1 &= 0.065 \\
    \mu &= 10/81 \\
    a_1^{\mr{SCAN}} &= 4.9479
\end{align}

\clearpage

\subsection{Correlation}

\begin{align}
    E_{\mr{c}}^\mathrm{r^2SCAN}[n_{\uparrow},n_{\downarrow}] &= \int n(\br) \epsilon_\mathrm{c}^\mathrm{r^2SCAN}(\br) d^3 r \\
	\epsilon_\mathrm{c}^\mathrm{r^2SCAN} &= \epsilon_\mathrm{c}^1 + f_{\mr{c}}(\bar\alpha)(\epsilon_\mathrm{c}^0 - \epsilon_\mathrm{c}^1) \\
	\epsilon_\mathrm{c}^1 &= \epsilon_\mathrm{c}^\mathrm{LSDA} + H_\mathrm{c}^1 \\
	\epsilon_\mathrm{c}^\mathrm{LDA0} &= -\frac{b_\mathrm{1c}}{1 + b_\mathrm{2c}\sqrt{r_s} + b_\mathrm{3c}r_s} \\
    f_{\mr{c}}(\bar\alpha) &= \begin{cases}
        \exp\left[-\frac{c_{1\mr{c}}^{\mr{SCAN}}\bar\alpha}{1 - \bar\alpha}\right] & \bar{\alpha} < 0 \\
        \sum_{i=0}^7c_{\mr{c}i}\bar\alpha^i & 0 \le \bar\alpha \leq 2.5 \\
        -d^{\text{SCAN}}_\mr{c}\exp\left[\frac{c^{\text{SCAN}}_{\mr{2c}}}{1-\bar\alpha}\right] & \bar\alpha > 2.5
        \end{cases}\label{eq:fc}\\
    \Delta f_\mathrm{c2} &=  \sum_{i=1}^7 ic_{\mr{c}i} \\
    d_s(\zeta) &= \frac{(1+\zeta)^{5/3} + (1-\zeta)^{5/3}}{2}\\
	H_\mathrm{c}^1 &= \gamma\phi^3\ln\left [1+w_1(1 - g(y, \Delta y))\right ] \\
	w_1 &= \exp\left[-\frac{\epsilon_\mathrm{c}^\mathrm{LSDA}}{\gamma\phi^3}\right ] - 1 \\
	g(y, \Delta y) &= \frac{1}{\left (1 + 4(y - \Delta y)\right )^{1/4}} \\
	y &= \frac{\beta(r_s)}{\gamma w_1}t^2\\
	\beta(r_s) &= \beta_{\mr{MB}}\frac{1 + 0.1 r_s}{1 + 0.1778 r_s} \\
    \Delta y &= \frac{\Delta f_{c2}}{27 \gamma d_s(\zeta) \phi^3 w_1} \left\{20r_s\left[\frac{\partial \varepsilon_c^{\text{LSDA0}}}{\partial r_s} - \frac{\partial \varepsilon_c^{\text{LSDA1}}}{\partial r_s} \right] - 45\eta[\varepsilon_c^{\text{LSDA0}} - \varepsilon_c^{\text{LSDA1}}] \right\} p \exp[-p^2/d_{p2}^4] \\
	\epsilon_\mathrm{c}^0 &= \left (\epsilon_\mathrm{c}^\mathrm{LDA0} + H_0\right ) G_\mathrm{c}(\zeta) \\
	H_0 &= b_\mathrm{1c}\ln\left [1 + w_0(1 - g_\infty(\zeta=0, s))\right ] \\
	G_c(\zeta) &= [1-2.3631(d_\mr{x}(\zeta) - 1)](1-\zeta^{12}) \\
	d_\mr{x}(\zeta) &= [(1+\zeta)^{4/3} + (1-\zeta)^{4/3}]/2 \\
	g_{\infty}(\zeta=0, s) &= \frac{1}{(1 + 4\chi_{\infty}s^2)^{1/4}}\\
	w_0 &= \exp\left [-\frac{\epsilon_\mathrm{c}^\mathrm{LDA0}}{b_\mathrm{1c}}\right ] - 1 \\
\end{align}

The values of the new constants appearing in the previous equations are the same as for exchange, $\eta = 0.001$ and $d_{p2}=0.361$.

Some constants appearing here are unchanged from either SCAN or rSCAN. Note that for the rSCAN interpolating polynomial we take the convention of indexing the coefficient by the power of $\alpha$ it multiplies, thus the vector $\bm{c}_c$ begins at $c_0$. These are:

\begin{align}
    \bm{c}_{c} &= (1, -0.64, -0.4352, -1.535685604549, 3.061560252175,\nonumber \\
    &  -1.915710236206, 0.516884468372, -0.051848879792) \\
    c^{\mr{SCAN}}_{1c} &= 0.64 \\
    c^{\mr{SCAN}}_{2c} &= 1.5 \\
    d^{\mr{SCAN}}_{c} &= 0.7 \\
    b_1 &= 0.0285764 \\
    b_2 &= 0.0889 \\
    b_3 &= 0.125541 \\
    \beta_{\mr{MB}} &\approx 0.066725 \\
    \chi_{\infty} &=\left(\frac{3\pi^2}{16}\right)^{2/3} \frac{\beta_{\mr{MB}}}{1.778\{ 0.9-3[3/(16\pi)]^{2/3}\}} \approx 0.128025
\end{align}

\clearpage

\section{Determining $\eta$ Regularization Parameter}

\begin{figure}[h]
    \centering
    \includegraphics{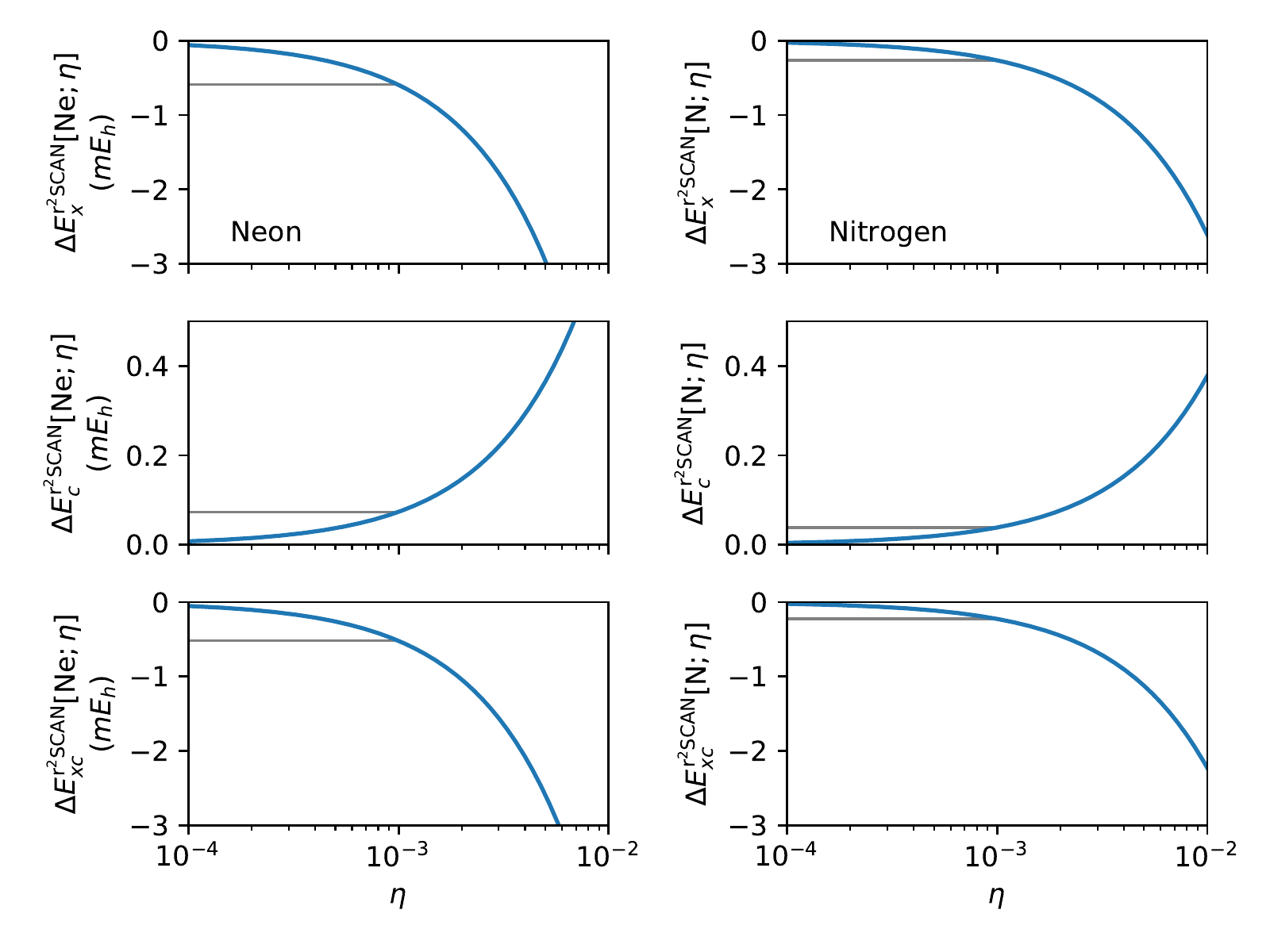}
    \caption{Change in energy, $\Delta E = E[\eta] - E[\eta = 0]$, as a function of the $\eta$ regularization parameter in r$^2$SCAN for exchange (top), correlation (middle), and exchange-correlation (lower) calculated from fixed Hartree--Fock Slater orbitals\cite{Clementi1974} for the neon and nitrogen atoms. All energies are monotonic functions of the regularisation parameter. The selected $\eta = 10^{-3}$ is shown by the horizontal line to the axis.}
    \label{fig:eta_atoms}
\end{figure}

\begin{figure}[h]
    \centering
    \includegraphics{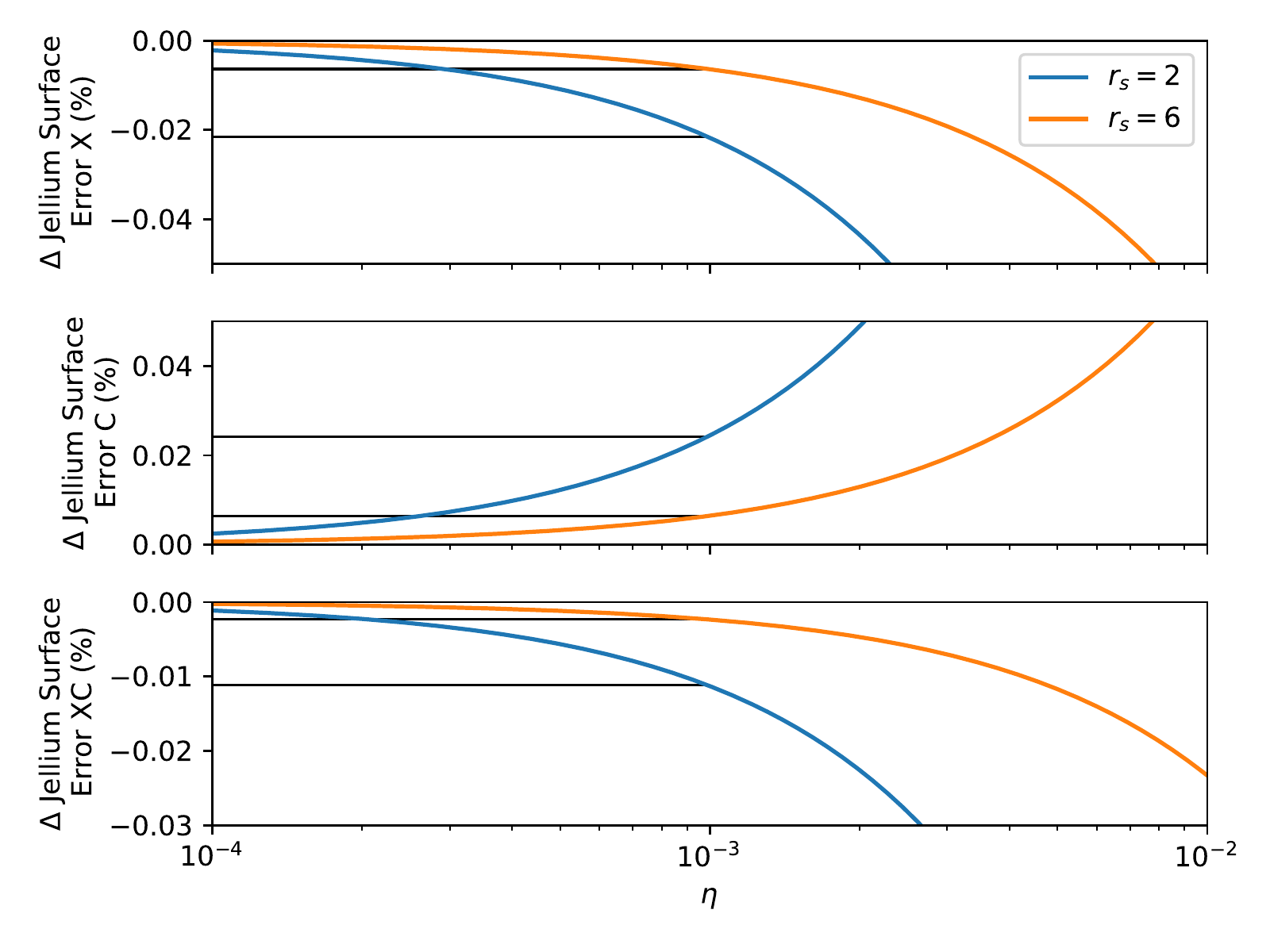}
    \caption{Change in percentage error for jellium surface energy ($r_s = 2, 6$), $\Delta E = E[\eta] - E[\eta = 0]$, as a function of the $\eta$ regularisation parameter in r$^2$SCAN for exchange (top), correlation (middle), and exchange-correlation (lower). All energies are monotonic functions of the regularization parameter. The selected $\eta = 10^{-3}$ is shown by the horizontal line to the axis.}
    \label{fig:eta_jelly}
\end{figure}

The effect of the $\eta$ regularization parameter on the r$^2$SCAN exchange and correlation functionals is illustrated in Figures \ref{fig:eta_atoms} and \ref{fig:eta_jelly}. For all systems studied the r$^2$SCAN total exchange, correlation, and exchange-correlation energies were monotonic functions of $\eta$, as expected. The value $\eta = 10^{-3}$ was arbitrarily selected as sufficiently large to ensure that $\bar{\alpha}$ remains finite at long ranges in practical calculations, while remaining small enough to have minimal impact on total energy.

\begin{figure}
    \centering
    \includegraphics[width=\textwidth]{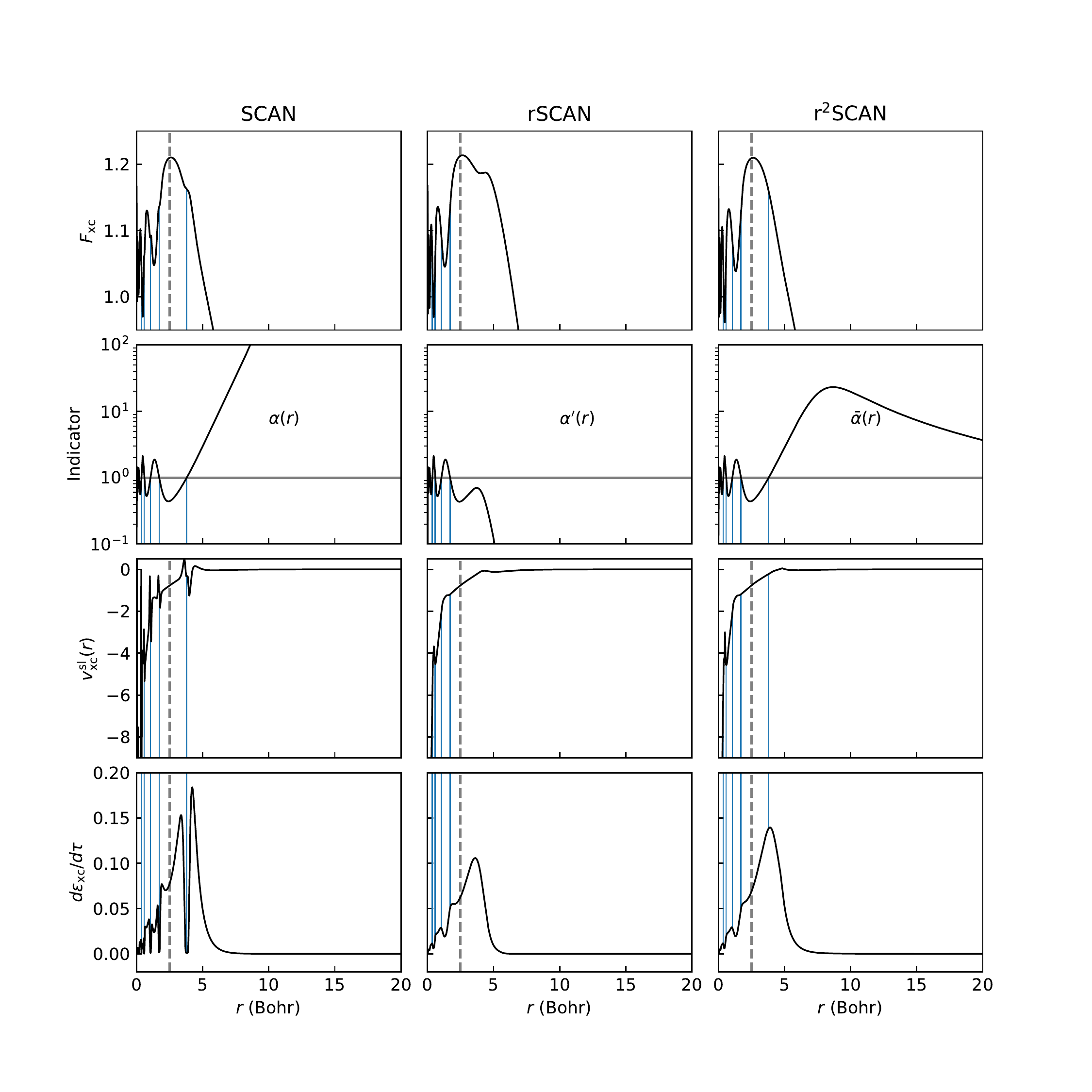}
    \caption{(Top) Exchange-correlation enhancement factors, (middle-upper, logarithmic scale) iso-orbital indicator $\alpha(r)$, $\alpha^\prime(r)$, or $\ba(r)$ as appropriate, (middle-lower) semi-local part of the exchange-correlation potential as in main text, and (bottom) derivative of exchange-correlation energy density with respect to kinetic energy density. Calculated for the xenon atom from accurate Hartree--Fock Slater orbitals \cite{Clementi1974} for the SCAN \cite{Sun2015}, rSCAN \cite{Bartok2019} and r$^2$SCAN functionals. The VASP \cite{Kresse1993,Kresse1994,Kresse1996,Kresse1996a} projector-augmented wave \cite{Joubert1999} pseudopotential cutoff radius (2.5 Bohr) is illustrated by a dashed vertical line. Solid vertical lines show where $\alpha=1$. This figure shows the same data as Figure 3 of the main text on a larger scale.}
    \label{fig:xe_lr}
\end{figure}

\begin{figure}
    \centering
    \includegraphics[width=\textwidth]{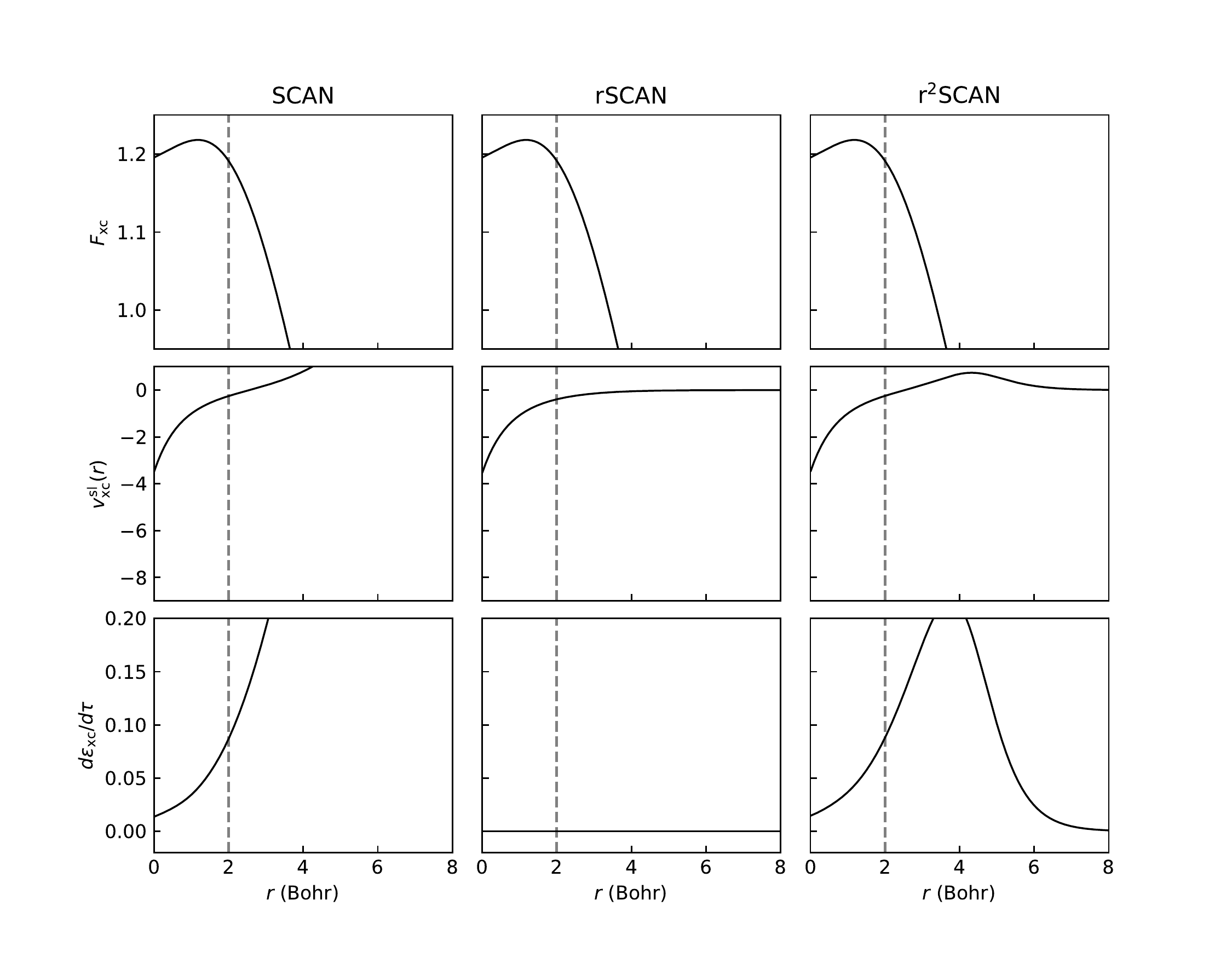}
    \caption{(Top) Exchange-correlation enhancement factors, (middle-upper, all zero for single orbital system) iso-orbital indicator $\alpha(r)$, $\alpha^\prime(r)$, or $\ba(r)$ as appropriate, (middle-lower) semi-local part of the exchange-correlation potential as in main text, and (bottom) derivative of exchange-correlation energy density with respect to kinetic energy density. Calculated for the helium atom from accurate Hartree--Fock Slater orbitals \cite{Clementi1974} for the SCAN \cite{Sun2015}, rSCAN \cite{Bartok2019} and r$^2$SCAN functionals. The VASP \cite{Kresse1993,Kresse1994,Kresse1996,Kresse1996a} projector-augmented wave \cite{Joubert1999} pseudopotential cutoff radius (2 Bohr) is illustrated by a dashed vertical line. Note the divergence of SCAN potential components.}
    \label{fig:he_pot}
\end{figure}

The exchange-correlation potential plots of Figure 3 in the main text are repeated here in Figure \ref{fig:xe_lr} to longer ranges and larger values of iso-orbital indicator, confirming that $\alpha$ diverges, while $\bar{\alpha}$ decays beyond $\sim8$ Bohr radii. The problematic divergence of SCAN potentials for single orbital systems is shown in Figure \ref{fig:he_pot}, where it is clear that the regularization introduced by r$^2$SCAN (and rSCAN) prevents this divergence in both multiplicative and $\tau$ dependent components.

\clearpage

\section{Reference Atomic Energies \label{sec:ref_at_en}}

We provide the total and exchange-correlation energies for the nitrogen and neon atoms from the common cc-pVTZ basis set \cite{Dunning1989,Woon1995} as references for testing implementations of the functionals. The largest ``reference'' level \textsc{Turbomole} grid was used {\color{Black}with increased radial grid density specified as \texttt{radsize 100} in the \texttt{control} file, as commonly suggested for the SCAN functional}.
\begin{longtable}{ll|rrrr}
    \caption{Reference self-consistent total atomic energies ($E$) \textbf{using only the exchange part of the functional}, and corresponding exchange energies ($E_{\mr{x}}$) in Hartree. Calculated using the cc-pVTZ basis set \cite{Dunning1989,Woon1995} and the ``reference'' level Turbomole grid.}  \label{tab:ref_at_dat_x} \\
    \endhead
         $\vphantom{\frac{1}{2}}$ & & SCAN & rSCAN & r$^2$SCAN \\
         \hline
        $\vphantom{\frac{1}{2}}$ \multirow{2}{*}{$E$} & N   & -54.40541174994 & -54.41398626268 & -54.39774875432 \\
        $\vphantom{\frac{1}{2}}$ & Ne  & -128.5891914977 & -128.6082003154 & -128.5698768302 \\
        \hline
        $\vphantom{\frac{1}{2}}$ \multirow{2}{*}{$E_{\mr{x}}$} & N  & -6.597139583375 & -6.604261530513 & -6.586631390511  \\
        $\vphantom{\frac{1}{2}}$  & Ne  & -12.143536468556 & -12.160698876055 & -12.11878636737 \\

\end{longtable}

\begin{longtable}{ll|rrrr}
    \caption{Reference self-consistent total atomic energies ($E$) and exchange-correlation energies ($E_{\mr{xc}}$) in Hartree calculated using the cc-pVTZ basis set \cite{Dunning1989,Woon1995} and the ``reference'' level Turbomole grid.}  \label{tab:ref_at_dat_xc} \\
     \endhead
          & & SCAN & rSCAN & r$^2$SCAN \\
         \hline
        \multirow{2}{*}{$E$} & N   & -54.58565736367 & -54.59430112747 & -54.57900797069 \\
         & Ne  & -128.9340794821 & -128.9541235246 & -128.9168416529 \\
        \hline
         \multirow{2}{*}{$E_{\mr{xc}}$} & N  & -6.782282570589 & -6.790099568534 & -6.773066377305  \\
          & Ne  & -12.49701065638 & -12.515767373154 & -12.474955269761 \\

\end{longtable}

\clearpage

\section{Test Set Data \label{sec:raw_dat}}

\begin{longtable}{l|rrrrrrr}
\caption{Deviation of TPSS atomisation energies (kcal/mol) for the G3 test set obtained with increasingly dense numerical grids. Mean error (ME) and Mean Absolute Error (MAE) are given at end of table.}\label{tab:G3_tpss}\\
\toprule
{} & 1 & 2 & 3 & 4 & 5 & 6 & 7 \\
\midrule
\endfirsthead
\caption{Deviation of TPSS atomisation energies (kcal/mol) for the G3 test set obtained with increasingly dense numerical grids. Mean error (ME) and Mean Absolute Error (MAE) are given at end of table.}\\
\toprule
{} & 1 & 2 & 3 & 4 & 5 & 6 & 7 \\
\midrule
\endhead
\midrule
\multicolumn{8}{r}{{Continued on next page}} \\
\midrule
\endfoot

\bottomrule
\endlastfoot
LiH                                                               &  -1.051 &  -1.089 &  -1.100 &  -1.098 &  -1.097 &  -1.097 &  -1.097 \\
BeH                                                               & -10.159 & -10.306 & -10.192 & -10.202 & -10.205 & -10.206 & -10.206 \\
CH                                                                &  -3.306 &  -3.405 &  -3.397 &  -3.398 &  -3.402 &  -3.402 &  -3.402 \\
CH\mol{2} (\excite{3}B\mol{1})                                    &  -8.104 &  -8.205 &  -8.204 &  -8.200 &  -8.208 &  -8.208 &  -8.207 \\
CH\mol{2} (\excite{1}A\mol{1})                                    &  -0.341 &  -0.513 &  -0.503 &  -0.503 &  -0.507 &  -0.507 &  -0.507 \\
CH\mol{3}                                                         &  -6.114 &  -6.322 &  -6.301 &  -6.300 &  -6.305 &  -6.305 &  -6.304 \\
Methane (CH\mol{4})                                               &  -4.373 &  -4.618 &  -4.614 &  -4.607 &  -4.612 &  -4.612 &  -4.612 \\
NH                                                                &  -6.650 &  -6.739 &  -6.672 &  -6.698 &  -6.698 &  -6.698 &  -6.698 \\
\end{longtable}

\clearpage

\begin{longtable}{l|rrrrrrr}
\caption{Deviation from reference ``exact'' values of SCAN atomization energies (kcal/mol) for the G3 test set obtained with increasingly dense numerical grids (numbered from 1 to 7). Mean error (ME) and Mean Absolute Error (MAE) are given at end of table.}\label{tab:G3_SCAN}\\
\toprule
{} & 1 & 2 & 3 & 4 & 5 & 6 & 7 \\
\midrule
\endfirsthead
\caption{Deviation from reference ``exact'' values of SCAN atomization energies (kcal/mol) for the G3 test set obtained with increasingly dense numerical grids (numbered from 1 to 7). Mean error (ME) and Mean Absolute Error (MAE) are given at end of table.}\\
\toprule
{} & 1 & 2 & 3 & 4 & 5 & 6 & 7 \\
\midrule
\endhead
\midrule
\multicolumn{8}{r}{{Continued on next page}} \\
\midrule
\endfoot

\bottomrule
\endlastfoot
LiH                                                               &   2.186 &   2.549 &   2.778 &   2.524 &   2.449 &   2.448 &   2.518 \\
BeH                                                               & -10.520 & -10.879 & -11.005 & -10.701 & -10.472 & -10.447 & -10.508 \\
CH                                                                &   1.438 &   1.265 &   0.127 &   1.596 &   2.317 &   2.277 &   1.914 \\
CH\mol{2} (\excite{3}B\mol{1})                                    &  -6.715 &  -6.243 &  -7.581 &  -6.325 &  -5.790 &  -5.911 &  -6.400 \\
CH\mol{2} (\excite{1}A\mol{1})                                    &   4.207 &   4.836 &   2.635 &   4.362 &   5.416 &   5.626 &   5.269 \\
CH\mol{3}                                                         &  -5.871 &  -5.316 &  -7.022 &  -5.373 &  -4.543 &  -4.490 &  -4.867 \\
Methane (CH\mol{4})                                               &   0.207 &  -0.173 &  -1.744 &  -0.707 &   0.398 &   0.746 &   0.570 \\
NH                                                                &  -2.770 &  -3.772 &  -1.302 &  -1.004 &  -1.294 &  -1.994 &  -1.230 \\
\end{longtable}

\clearpage

\begin{longtable}{l|rrrrrrr}
\caption{Deviation from reference ``exact'' values of rSCAN atomization energies (kcal/mol) for the G3 test set obtained with increasingly dense numerical grids (numbered from 1 to 7). Mean error (ME) and Mean Absolute Error (MAE) are given at end of table.}\label{tab:G3_rSCAN}\\
\toprule
{} & 1 & 2 & 3 & 4 & 5 & 6 & 7 \\
\midrule
\endfirsthead
\caption{Deviation from reference ``exact'' values of rSCAN atomization energies (kcal/mol) for the G3 test set obtained with increasingly dense numerical grids (numbered from 1 to 7). Mean error (ME) and Mean Absolute Error (MAE) are given at end of table.}\\
\toprule
{} & 1 & 2 & 3 & 4 & 5 & 6 & 7 \\
\midrule
\endhead
\midrule
\multicolumn{8}{r}{{Continued on next page}} \\
\midrule
\endfoot

\bottomrule
\endlastfoot
LiH                                                               &    2.022 &    2.093 &    2.080 &    2.144 &    2.139 &    2.125 &    2.130 \\
BeH                                                               &  -11.229 &  -11.179 &  -10.945 &  -11.018 &  -11.011 &  -10.996 &  -11.004 \\
CH                                                                &    1.522 &    1.462 &    1.516 &    1.492 &    1.388 &    1.407 &    1.427 \\
CH\mol{2} (\excite{3}B\mol{1})                                    &   -8.619 &   -8.969 &   -8.713 &   -8.713 &   -8.797 &   -8.788 &   -8.767 \\
CH\mol{2} (\excite{1}A\mol{1})                                    &    3.672 &    3.601 &    3.784 &    3.737 &    3.597 &    3.612 &    3.645 \\
CH\mol{3}                                                         &   -7.133 &   -7.526 &   -7.234 &   -7.322 &   -7.427 &   -7.407 &   -7.384 \\
Methane (CH\mol{4})                                               &   -3.787 &   -3.666 &   -3.561 &   -3.535 &   -3.678 &   -3.666 &   -3.633 \\
NH                                                                &   -1.620 &   -1.739 &   -1.829 &   -2.047 &   -1.987 &   -1.930 &   -1.959 \\
\end{longtable}

\clearpage


\begin{longtable}{l|rrrrrrr}
\caption{Deviation from reference ``exact'' values of r${}^2$SCAN atomization energies (kcal/mol) for the G3 test set obtained with increasingly dense numerical grids (numbered from 1 to 7). Mean error (ME) and Mean Absolute Error (MAE) are given at end of table.}\label{tab:G3_r2SCAN}\\
\toprule
{} & 1 & 2 & 3 & 4 & 5 & 6 & 7 \\
\midrule
\endfirsthead
\caption{Deviation from reference ``exact'' values of r${}^2$SCAN atomization energies (kcal/mol) for the G3 test set obtained with increasingly dense numerical grids (numbered from 1 to 7). Mean error (ME) and Mean Absolute Error (MAE) are given at end of table.}\\
\toprule
{} & 1 & 2 & 3 & 4 & 5 & 6 & 7 \\
\midrule
\endhead
\midrule
\multicolumn{8}{r}{{Continued on next page}} \\
\midrule
\endfoot

\bottomrule
\endlastfoot
LiH                                                               &     2.456 &     2.527 &     2.485 &     2.554 &     2.587 &     2.584 &     2.576 \\
BeH                                                               &   -10.429 &   -10.482 &   -10.241 &   -10.284 &   -10.263 &   -10.249 &   -10.258 \\
CH                                                                &     2.636 &     2.473 &     2.574 &     2.526 &     2.434 &     2.458 &     2.473 \\
CH\mol{2} (\excite{3}B\mol{1})                                    &    -6.153 &    -6.572 &    -6.330 &    -6.329 &    -6.401 &    -6.386 &    -6.371 \\
CH\mol{2} (\excite{1}A\mol{1})                                    &     5.975 &     5.749 &     5.977 &     5.924 &     5.788 &     5.810 &     5.838 \\
CH\mol{3}                                                         &    -3.492 &    -4.114 &    -3.772 &    -3.862 &    -3.961 &    -3.934 &    -3.917 \\
Methane (CH\mol{4})                                               &     1.087 &     0.687 &     1.033 &     1.002 &     0.860 &     0.876 &     0.904 \\
NH                                                                &    -0.300 &    -0.680 &    -0.718 &    -0.906 &    -0.854 &    -0.805 &    -0.833 \\
\end{longtable}

\clearpage

\begin{longtable}{l|rrrr}
\caption{Error in reaction barrier heights (kcal/mol) for the BH76 test set. Mean error (ME) and Mean Absolute Error (MAE) are given at end of table.}\label{tab:full_BH76}\\
\toprule
{} &  TPSS &  SCAN &  rSCAN &  r${}^2$SCAN \\
\midrule
\endfirsthead
\caption{Error in reaction barrier heights (kcal/mol) for the BH76 test set. Mean error (ME) and Mean Absolute Error (MAE) are given at end of table.}\\
\toprule
{} &  TPSS &  SCAN &  rSCAN &  r${}^2$SCAN \\
\midrule
\endhead
\midrule
\multicolumn{5}{r}{{Continued on next page}} \\
\midrule
\endfoot

\bottomrule
\endlastfoot
0   &  -10.214 &   -8.791 &    -8.273 &     -7.781 \\
1   &  -23.136 &  -18.390 &   -21.305 &    -20.121 \\
2   &  -15.505 &  -13.497 &   -13.512 &    -13.044 \\
3   &  -15.505 &  -13.497 &   -13.512 &    -13.044 \\
4   &  -10.490 &   -8.797 &    -9.233 &     -8.601 \\
5   &  -10.490 &   -8.797 &    -9.233 &     -8.601 \\
6   &  -13.075 &  -10.468 &   -11.605 &    -11.038 \\
7   &  -14.959 &  -10.804 &   -10.046 &    -10.149 \\
\end{longtable}

\clearpage

\begin{longtable}{l|rrrr}
\caption{Error in dissociation energies (kcal/mol) for the S22 test set. Mean error (ME) and Mean Absolute Error (MAE) are given at end of table.}\label{tab:full_S22}\\
\toprule
{} &  TPSS & SCAN & rSCAN &  r${}^2$SCAN \\
\midrule
\endhead
\midrule
\multicolumn{5}{r}{{Continued on next page}} \\
\midrule
\endfoot

\bottomrule
\endlastfoot
0   &   -0.889 &   -0.032 &    -0.259 &     -0.236 \\
1   &   -0.532 &    0.383 &     0.051 &      0.135 \\
2   &   -0.915 &    2.157 &     1.332 &      1.248 \\
3   &   -1.995 &    0.478 &    -0.181 &     -0.181 \\
4   &   -2.728 &    0.025 &    -0.573 &     -0.640 \\
5   &   -2.472 &    0.031 &    -0.434 &     -0.547 \\
6   &   -3.268 &   -0.502 &    -1.045 &     -1.108 \\
7   &   -0.652 &   -0.176 &    -0.298 &     -0.238 \\
8   &   -1.845 &   -0.450 &    -0.618 &     -0.614 \\
9   &   -1.890 &   -0.498 &    -0.862 &     -0.660 \\
10  &   -5.488 &   -1.375 &    -2.431 &     -1.824 \\
11  &   -6.015 &   -1.370 &    -2.379 &     -1.868 \\
12  &   -8.413 &   -1.366 &    -2.605 &     -2.000 \\
13  &   -8.215 &   -2.335 &    -3.678 &     -2.962 \\
14  &  -12.093 &   -2.553 &    -4.153 &     -3.500 \\
15  &   -0.588 &   -0.076 &    -0.276 &     -0.191 \\
16  &   -1.800 &    0.075 &    -0.421 &     -0.108 \\
17  &   -1.911 &   -0.235 &    -0.689 &     -0.406 \\
18  &   -2.183 &   -0.348 &    -0.797 &     -0.580 \\
19  &   -3.167 &   -1.012 &    -1.494 &     -1.262 \\
20  &   -4.351 &   -1.344 &    -1.950 &     -1.619 \\
21  &   -4.053 &   -1.017 &    -1.599 &     -1.454 \\
\hline
ME  &   -3.430 &   -0.524 &    -1.153 &     -0.937 \\
MAE &    3.430 &    0.798 &     1.273 &      1.057 \\
\end{longtable}

\clearpage

\begin{longtable}{l|rrrr}
\caption{Error in lattice constants ($\text{\AA}$) for the LC20 test set. Mean error (ME) and Mean Absolute Error (MAE) are given at end of table.}\label{tab:full_LC20}\\
\toprule
{} & TPSS & SCAN &  rSCAN & r${}^2$SCAN \\
\midrule
\endhead
\midrule
\multicolumn{5}{r}{{Continued on next page}} \\
\midrule
\endfoot

\bottomrule
\endlastfoot
Ag & 0.019 & 0.012 & 0.028 & 0.034 \\
Al & -0.011 & -0.012 & -0.027 & -0.032 \\
Ba & 0.007 & 0.049 & 0.100 & 0.076 \\
C & 0.014 & -0.004 & -0.001 & 0.005 \\
Ca & -0.032 & -0.009 & 0.017 & 0.018 \\
Cu & -0.016 & -0.030 & -0.023 & -0.020 \\
GaAs & 0.075 & 0.020 & 0.031 & 0.029 \\
Ge & 0.083 & 0.029 & 0.043 & 0.039 \\
Li & -0.014 & 0.011 & 0.021 & 0.016 \\
LiCl & 0.040 & 0.016 & 0.021 & 0.034 \\
LiF & 0.070 & 0.004 & 0.008 & 0.021 \\
MgO & 0.056 & 0.018 & 0.019 & 0.027 \\
NaCl & 0.129 & 0.010 & 0.025 & 0.036 \\
NaF & 0.133 & 0.006 & 0.015 & 0.028 \\
Na & 0.032 & -0.007 & 0.024 & 0.004 \\
Pd & 0.025 & 0.016 & 0.026 & 0.032 \\
Rh & 0.011 & -0.005 & 0.006 & 0.008 \\
Si & 0.033 & 0.005 & 0.012 & 0.018 \\
SiC & 0.019 & 0.002 & -0.001 & 0.006 \\
Sr & -0.014 & 0.039 & 0.064 & 0.056 \\
ME & 0.033 & 0.009 & 0.020 & 0.022 \\
MAE & 0.041 & 0.015 & 0.025 & 0.027 \\
\end{longtable}



\end{document}